\documentstyle[preprint,prb,aps]{revtex}

\begin{document}

\title{SURPRISES IN THE ORBITAL MAGNETIC MOMENT AND
$g$-FACTOR OF THE DYNAMIC JAHN-TELLER ION C$_{60}^-$}

\author{ Erio
Tosatti$^{1,2,3}$\thanks{Email: tosatti@sissa.it},
Nicola Manini$^{1,2,4}$\thanks{Email: manini@esrf.fr},and Olle
Gunnarsson$^{5}$\thanks{Email: gunnar@radix3.mpi-stuttgart.mpg.de}}

\address{$^1$ S.I.S.S.A., Via Beirut 4, I-34013 Trieste, Italia}

\address{$^2$ Istituto Nazionale di Fisica della Materia (INFM)}

\address{$^3$ I.C.T.P., P.O. Box 586, I-34014 Trieste, Italia}

\address{$^4$ E.S.R.F., B.P. 220, F 38043 Grenoble Cedex, France}

\address{$^5$ M.P.I. f\"ur Festk\"orperforschung, D-70569 Stuttgart, F.R.G.}

\maketitle

\begin{abstract}
We calculate the magnetic susceptibility and $g$-factor of the isolated
C$_{60}^{-}$ ion at zero temperature, with a proper treatment of the
dynamical Jahn-Teller effect, and of the associated orbital angular
momentum, Ham-reduced gyromagnetic ratio, and molecular spin-orbit
coupling. A number of surprises emerge.  First, the predicted molecular
spin-orbit splitting is two orders of magnitude smaller than in
the bare carbon atom, due to the large radius of curvature of the molecule.
Second, this reduced spin-orbit splitting is comparable
to Zeeman energies, for instance, in X-band EPR at 3.39KGauss, and a
field dependence of the $g$-factor is predicted.  Third, the
orbital gyromagnetic factor is strongly reduced by vibron coupling, and so
therefore are the effective weak-field $g$-factors of all low-lying
states. In particular, the ground state doublet of C$_{60}^{-}$ is predicted to
show a negative $g$-factor of $\sim -0.1$.
\end{abstract}


\section{Introduction}

A neutral, isolated fullerene molecule is an eminently stable and
symmetrical system. The fullerene ions (either negative or positive) along
with the electronically excited neutral molecule, particularly the
long-lived triplet exciton, undergo instead Jahn-Teller (JT) distortions.
The negative ion and the triplet exciton, respectively with $t_{1u}$ and
$t_{1g}$ symmetry, will distort according to a linear combination of the
eight $H_g$ molecular modes. The JT distortion of the positive ion, with
$h_u$ symmetry, involves also the six $G_g$ modes in addition to the
eight $H_g$ ones.  Although accurate numerical values of all couplings are
not yet available in all cases, the static JT energy gains are believed to
be roughly in the order of 0.1eV. This value is comparable with the typical
vibrational frequency, and the coupling is generally of intermediate
strength. Several descriptions of the static JT effect in fullerene ions
can be found in the literature.\cite {Negri,vzr,Schl,Johnson}

As pointed out more recently, however, a static JT description, where the
pseudorotational motion of the carbon nuclei is treated classically, and
then quantized separately in the Born-Oppenheimer approximation is, at
least in the isolated molecule, and at zero temperature
fundamentally inadequate. In other words, the fullerene ions are expected
to be genuine {\it dynamical }Jahn-Teller (DJT) systems, where different
but equivalent distorted configurations, (forming the usual static JT
manifold\cite{Ceulemans}), are not independent of one another, but are in
fact connected by nonzero transition amplitudes\cite{Wang}. This in
turn requires
giving up Born-Oppenheimer, and fully quantizing electronic and ionic
motions together, which is the essence of DJT.

The physical understanding of DJT in C$_{60}$ ions is greatly eased by
initially assuming the strong-coupling limit. In this limit, as it turns
out, a {\em modified} Born-Oppenheimer approximation can again be
recovered, provided a suitable gauge field, reflecting the electronic {\em
Berry phase},\cite{Berry} is added to the nuclear motion. This situation,
discussed originally for the triatomic molecule,\cite{Delacretaz} and
subsequently for other JT systems\cite{o'brien} has been
recently the object of a close scrutiny in fullerene, especially in the
negative ions\cite{AMT,MTA,Ihm} and, to a lesser degree, in the positive
ion.\cite{delos} It is found in particular that, if treated in the
strong-coupling limit, the odd-charged fullerene ions, in particular the
singly-charged C$_{60}^{-}$\cite{AMT,Reno} and C$_{60}^{+}$,
\cite{delos,ChoB95}, the Hund's rule {\em triplet} ground state of
doubly charged
C$_{60}^{2-}$, as well as the neutral $t_{1g}$ triplet
exciton,\cite{triplet:note}, must possess this kind of Berry phase. By
contrast, the Berry phases cancel out in the {\em singlet} configuration of
even-charged ions, such as C$_{60}^{2-}$, and C$_{60}^{2+}$. Although, as
stated above, the true electron-vibron coupling in C$_{60}$ is in reality
only of intermediate strength, the presence or absence of a Berry phase in
the strong-coupling limit DJT implies a number of physical consequences,
which persist at realistic couplings, and whose importance has been
discussed in detail elsewhere.  Properties affected include basic
ground-state features such as symmetry\cite{ham} and energy,\cite{MTA,Reno}
spectral features (including characteristic splittings of the lowest vibron
excitations with skipping of even angular
momenta\cite{AMT,MTA,Reno,Gunnarsson}), scattering anomalies such as
suppression of ordinary s-wave attachment of low-energy
electrons,\cite{attachment} and the prediction of orbitally-related
electron pairing phenomena in idealized molecular metal lattices with weak
electron hopping between molecules.\cite{MTD,SAMTP,SMPT,Fabrizio}

For these and other reasons it seems important to understand completely and
quantitatively the DJT effect of single fullerene ions, experimentally as
well as theoretically. To this date, however, and in spite of a large
amount of data collected on fullerene and especially on fulleride systems,
there is still frustratingly little {\it direct} evidence that the JT
effect in fullerene ions is, to start with, really dynamical, as theory
predicts.

In the solid state, for example, Raman data on metallic fullerides such as
K$_3$C$_{60}$\cite{Zhou2} fail to show the characteristic vibron splittings
expected for the isolated C$_{60}^{3-}$ ions.\cite{MTA,Reno} This probably
means that in true metallic fullerides the molecular JT effect may be
profoundly affected and modified by the large crystal fields, as well as by
the strong, rather than weak, intermolecular electron hopping. Even in
nonmetallic fullerides like K$_4$C$_{60}$, where the insulating behavior is
almost certainly due to a molecular JT effect,\cite{MTA,Kerkoud} it is
presently not at all clear whether the quantum dynamical features are
present or suppressed. Fullerene ions have also been widely studied in
solutions\cite{Dubois} and in solid ionic salts\cite{kato} but again no
specific DJT signature has been pinpointed, so far.

The main target remains therefore {\it gas-phase fullerene}: the DJT
signatures should be unmistakable in ions of either sign, and in
triplet-excited neutral molecules, particularly if in the future Raman
excitations could be studied.\cite{Reno} What is available so far are
essentially only gas-phase photoemission spectra of
C$_{60}^{-}$,\cite{Gunnarsson} and of C$_{60}$.\cite{bruhwiler}
Encouragingly, the former are fit very well indeed by a DJT
theory.\cite{Gunnarsson} Nonetheless, this can only be considered as
indirect evidence. The positive ions results have not yet been analyzed,
although the appropriate DJT theory has been formulated.\cite{delos}

We therefore wish to consider here other properties of the fullerene ions,
among those crucially affected by the DJT effect, which could at least
in principle be
accessed either in gas-phase, or in ideally inert matrices, or in
suitable salts with especially small crystal-field effects. One such
quantity is precisely the molecular magnetic moment. The magnetic moment of
a {\em static} JT molecule is strictly the spin moment. The orbital
degeneracy is removed, so long as static-JT energies are, as in the case of
C$_{60}$, sufficiently large. The magnetic moment in a {\em dynamic}-JT
molecule, conversely, is a compound of spin and orbital moments, since here
the quantum effects fully restore the original orbital symmetry.\cite{ham}
The calculation of the magnetic moment and effective $g$-factor of
fullerene ions in their DJT ground state is precisely the subject of this
work.

Let us consider for a start the orbital magnetic moment of a molecule.
Qualitatively speaking, the proportionality factor between magnetic moment
and the mechanical angular momentum can be thought as some {\it effective}
Bohr magneton $e\hbar /2m^{*}$, where $m^{*}$ is the mass of the orbiting
electron ($e$ is the electron charge and S.I. units are used
throughout). In an atom, $m^{*}=m_e$, the free electron mass. In a DJT
molecule, however, orbital electron motion involves nuclear motion as well,
since electronic and vibrational modes are entangled. Therefore, we expect
$m^{*}>m_e$, with a corresponding reduction of the orbital magnetic
moment. This is a classic example of the so-called ``Ham reduction
factor'',\cite{ham65} well-known in DJT systems.\cite{Englman} The
consequence of orbital reduction is that, while the quantum of mechanical
angular momentum is of course universal and equal to $\hbar$, that of the
magnetic moment for a DJT molecule is {\em not universal}.

Below, we will calculate quantitatively the orbital magnetic moment for
C$_{60}^{-}$. Moreover, since the orbital moment is not easily accessible
experimentally, while the total magnetic moment is commonly measured,
spin-orbit coupling will have to be introduced, to determine the correct
composition of the (DJT-reduced) orbital moment, and of the spin moment.
The spin-orbit coupling within the $t_{1u}$ orbital of C$_{60}$ is
quantitatively unknown. Here we calculate it, and find some surprises.
First, the calculated molecular spin-orbit splitting is very small, roughly
one hundredth than in the bare carbon atom. This is related to the larger
radius of curvature of the molecular orbit.  Second, this reduced
spin-orbit splitting is now wholly comparable to typical Zeeman energies,
for instance,in X-band EPR at 0.339T. Hence, a field dependence of the
low-temperature susceptibility and of apparent $g$-factors is predicted,
at least in
an idealized gas-phase EPR experiment.  Third, we find that the orbital
gyromagnetic factor is strongly reduced by vibron coupling, and so
therefore are the effective weak-field $g$-factors for all the low-lying
states. The ground state $g$-factor, in particular, is predicted to be
slightly {\em negative}, about -0.1.

\section{The model}
\label{model:sect}

The basic model Hamiltonian we consider has the following standard
structure:
\begin{equation}
\label{Hschemat:eq}H=H^0+H^{e-v}+H^{so}+H^B\ .
\end{equation}
The JT part $H^0+H^{e-v}$ has been introduced and discussed in previous
papers.\cite{AMT,MTA,Gunnarsson} We report here the basic version for the
coupling to a single $H_g$ vibrational (quadrupolar) mode:
\begin{eqnarray}
H^0 & = & \hbar \omega \sum_{m=-2}^2(b_m^{\dagger }b_m+\frac 12)+(\epsilon
-\mu) \sum_{m=-1}^1\sum_{\sigma =\uparrow,\downarrow }
c_{m,\sigma}^{\dagger }c_{m,\sigma }\ , \nonumber\\
\label{Hesplic:eq}
H^{e-v} & = & g \frac{\sqrt{3}}2\hbar \omega \sum_{m_1,m_2,\sigma}
(-1)^{m_2}<1,m_1;1,m_2\mid 2,m_1+m_2> \\
&  & \ \ \ \times \left[ b_{m_1+m_2}^{\dagger}
+(-1)^{m_1+m_2}b_{-m_1-m_2}\right]
c_{m_1,\sigma }^{\dagger}c_{-m_2,\sigma}\
+\ ... \nonumber
\end{eqnarray}
$H^0$ describes the free (uncoupled) electrons and the fivefold-degenerate
vibration of frequency $\omega$. $H^{e-v}$ introduces a (rotationally
invariant) standard linear coupling between the electronic state and the
vibrational mode. The dimensionless linear coupling parameter is indicated
as $g$ (not to be confused with the magnetic factors $g_L$). Here we
neglect higher-order terms in the boson operators, indicated with
continuation dots.

Orbital currents are associated with the partly filled $t_{1u}$ level,
which is known to derive essentially from a superatomic $L=5$ orbital of
C$_{60}$ as a whole.\cite{Savina,Troullier} These orbital currents give
rise to a magnetic moment, which we now wish to calculate.

Electron spin also contributes to the total magnetic moment. Although
uninfluenced by JT coupling (the Hamiltonian (\ref{Hesplic:eq}) conserves
spin), spin is coupled to the orbital motion {\it via} spin-orbit coupling
$H^{so}=\lambda ({\vec L}\cdot {\vec S})$. The general Hamiltonian finally
includes Zeeman coupling to an external magnetic field $B$ along the $z$
axis
\begin{equation}
\label{HZeeman1:eqn}H^B=-\mu_B B (g_LL_z+g_SS_z)
\end{equation}
where $\mu_B=e\hbar /2m$.  For generic $g_L$ and $g_S$ factors, the
appropriate value for the $g_J$ factor of the spin-orbit coupled state
($\left| L-S\right| \leq J\leq \left| L+S\right| $) is\cite{Abragam}
\begin{equation}
\label{CoupledGfac:eqn}
g_J=\frac{g_L+g_S}2+\frac{L(L+1)-S(S+1)}{2J(J+1)}\left(g_L-g_S\right) \ .
\end{equation}
For C$_{60}^{-}$, $L=1$, $S=\frac 12$, $J=\frac 12$,\ $\frac 32$, and
$g_S=2.0023$ as appropriate for a free spin. Thus, in order to obtain the
$g_J $ factors of the individual spin-orbit split states $J=\frac 12$ and
$J=\frac 32$, we only need to calculate the value of $g_L$.

Two main phenomena should affect the value of the orbital $g_L$-factor: the
${\cal D}^{(L=5)}$ parentage of the $t_{1u}$ (${\cal D}^{(L=1)}$) state,
and the DJT coupling with the vibrons.

An old paper by Cohan\cite{Cohan} provides tables with the icosahedral
decomposition of spherical states up to $L=15$, as expansion coefficients
on an unnormalized and real basis
\begin{equation}
\label{UnnormSpHar:eqn}
\widetilde{Y}_{L,|M|}^{C/S}=\left[ \frac{4\pi }{2L+1}\
\frac{\left(L+|M|\right) !}{\left(L-|M|\right) !}\right]^{\frac12}
\frac{Y_{L,M}\pm Y_{L,-M}}{2(-1)^{\frac 34\pm \frac 14}}
\end{equation}
The tabulated wave functions which transform as $t_{1u}$ are:
\begin{eqnarray}
\psi_0&\propto& 2160\ \widetilde{Y}_{5,0}^C\ +\ \widetilde{Y}_{5,5}^C
\nonumber\\
\psi_{C/S}&\propto& 72\ \widetilde{Y}_{5,1}^{C/S}\ \mp \
\widetilde{Y}_{5,4}^{C/S}\ ,
\label{PsiC/S:eqn}
\end{eqnarray}
which can be normalized to obtain
\begin{eqnarray}
\psi _0&=&
	{\cal C}\ \left[ 2160\ Y_{5,0}\ +\frac{\sqrt{10!}}2
	\left(Y_{5,5}+Y_{5,-5}\right) \right]
=	\frac 6{\sqrt{50}}Y_{5,0}+
	\sqrt{\frac 7{50}}\left(Y_{5,5}+Y_{5,-5}\right) \nonumber\\
\label{PSIC/S:eqn}
\psi _{C/S}&=&{\cal C}\ \left[ 72\sqrt{\frac{6!}{4!}}\
	\left(Y_{5,1}\pm Y_{5,-1}\right) \ \mp \sqrt{\frac{9!}{1!}}
	\left(Y_{5,4}\pm Y_{5,-4}\right) \right] =\\
& &	\frac 1{\sqrt{2}~(-1)^{\frac34 \pm \frac14}}\left[ \sqrt{\frac 3{10}}\
	\left(Y_{5,1}\pm Y_{5,-1}\right) \ \mp \sqrt{\frac 7{10}}\left(
	Y_{5,4}\pm Y_{5,-4}\right) \right] \ , \nonumber
\end{eqnarray}
yielding
\begin{equation}
\label{PSIPM:eqn}
\psi _{\mp 1}=
\frac{\psi _C\pm i\psi _S}{\sqrt{2}}=
\sqrt{\frac 3{10}}\ Y_{5,\mp 1}-\sqrt{\frac 7{10}}Y_{5,\pm 4}\ .
\end{equation}

On the $\left\{ \psi _M\right\} _{M=-1,0,1}$ basis of $t_{1u}$, the orbital
Zeeman coupling with the external magnetic field is diagonal:
\begin{equation}
\label{gEffPar:eqn}\left\langle \psi _M\right| H_L^B\left| \psi_{M^{\prime
}}\right\rangle =-g_1\mu_B\left[ \frac 3{10}(-1)+\frac 7{10}\cdot
(4)\right] B\ M\ \delta_{MM^{\prime }}=-\frac 52g_1\mu_B B\ M\
\delta_{MM^{\prime }}
\end{equation}
This formula implicitly defines an effective orbital $\widetilde{g}_1$
factor $\frac 52g_1$, ($\alpha = \frac 52$, in the language of Ref.\
\cite{Abragam}). It is convenient to define for the $t_{1u}$ orbital an
effective angular momentum $\vec {\widetilde{L}}$, whose $z$-component has
values -1,0,1 on the $\left\{ \psi _M\right\}_{M=-1,0,1}$ basis, in terms
of which the orbital Zeeman interaction (\ref{HZeeman1:eqn}) is rewritten
\begin{equation}
\label{HZeeman2:eqn}
H_L^B=-\widetilde{g}_L \mu_B B\widetilde{L}_z=
-\alpha g_L \mu_B B\widetilde{L}_z
\end{equation}
It is clear that the enhancement is due to the $|M|=4$ component in the the
$t_{1u}$ wave function, which is really $L=5$, but is regarded formally as
an effective $\widetilde{L}=1$ state. Reference\ \cite{Troullier} also
provides the explicit spherical parentage of the LUMO orbital, now in a
solid-state environment.  By computing the spectrum of $L_z$ on the basis
provided in that work, we obtain a somewhat smaller value for $\alpha $,
namely $1.86$. However, for gas-phase C$_{60}^{-}$ the group-theoretical
value for a strictly $L=5$ parentage, $\alpha =2.5$, is probably more
accurate.

We now concentrate on the second effect, namely that of the coupling of the
electronic state with the vibrons. For clarity we start considering a
single $H_g$ vibron coupled to the $t_{1u}$ level, as in
Eq.~(\ref{Hschemat:eq}). We represent the $L_z$ operator in second
quantization as $\sum_\sigma \left(c_{1\sigma }^{\dagger }c_{1\sigma
}-c_{-1\sigma }^{\dagger }c_{-1\sigma }\right)$ (same notation as in
Eq.~(\ref{Hesplic:eq})) and we measure the magnetic energy $\mu_B B$ in units
of the energy scale of the vibron, $\hbar \omega $, and indicate it with
${\cal B}\equiv \frac{\mu_B B}{\hbar\omega}$.

As in Ref.\ \cite{AMT}, it is instructive to treat first the
weak-JT-coupling limit. We solve the quantum problem in perturbation theory
to second order in the e-v coupling parameter $g$, this time including
$H^B$ (which is diagonal on the basis $\left| \psi_M\right\rangle$) in the
unperturbed Hamiltonian.  Thus we consider $H=\left(H^0+H^B\right)
+H^{e-v}$, and apply nondegenerate perturbation theory to second order
within the threefold space of the $t_{1u} $ level (\ref{PSIC/S:eqn}). The
second-order energy shift caused by $H^{e-v}$ to level $|\psi _M>$
($M=-1,0,1$, unperturbed energy $\left(\frac 52-\widetilde{g}_1M\cdot {\cal
B}\right) \hbar \omega $) is:
\begin{equation}
\label{5.3}\Delta _M^{(2)}=\left\langle \psi _M\right| H^{e-v}\frac 1{\left(
\frac 52-\widetilde{g}_1M\cdot {\cal B}\right)
\hbar \omega {-}\left(H^0+H^B\right) }H^{e-v}\left| \psi _M\right\rangle
\end{equation}
while off diagonal terms $\Delta _{MM^{\prime }}^{(2)}$ vanish since
$H^{e-v}$ is rotationally invariant. This shift can be rewritten as
\begin{equation}
\label{shift}\Delta _M^{(2)}=-\frac 34 g^2\hbar \omega
\sum_{m=-1,0,1}\frac{(\langle 1,m;1,-M|2,m-M\rangle)^2}
{1+\widetilde{g}_1{\cal B}(M-m)}
\end{equation}
By substituting the Clebsch-Gordan coefficients, and carrying out the sum
over $m$, for each fixed value of $M$, we get
\begin{eqnarray}
\label{shift0}
\Delta _0^{(2)}&=&-\frac 34g^2\hbar \omega \left[ \frac 23+ \frac 12\left(
\frac 1{1-\widetilde{g}_1{\cal B}}+\frac 1{1+\widetilde{g}_1 {\cal
B}}\right) \right] =-\frac 34g^2\hbar \omega \left[ \frac 53+{\cal O}({\cal
B} ^2)\right] \\
\Delta _{\pm 1}^{(2)}&=&-\frac 34g^2\hbar \omega \left[ \frac
16+\frac 12\frac 1{1\pm \widetilde{g}_1{\cal B}}+
\frac 1{1\pm 2\widetilde{g}_1{\cal B}}\right] =
-\frac 34g^2\hbar \omega \left[ \frac 53\mp \frac 52
\widetilde{g}_1{\cal B}+{\cal O}({\cal B}^2)\right]
\nonumber
\end{eqnarray}
The weak-field ${\cal B}$-expansion, is done here under the customary
assumption that the magnetic energy is the smallest energy scale in the
problem (we will return to this point later, however). The final result for
the energy to first order in $\cal B$ of the three $t_{1u}$- derived levels
is finally
\begin{equation}
\label{EsecOrd:eqn}E_M^{(2)}=\left(\frac 52-\frac 54 g^2 \right) \hbar \omega
-M\left(1-\frac{15}8g^2\right) \widetilde{g}_1{\cal B\ }\hbar \omega \ .
\end{equation}
The result of e-v coupling is a reduction of both zero-point energy
($-\frac 54 g^2\hbar \omega$), and magnetic moment.
By identification we obtain
\begin{equation}
\label{Geff1}
g_1^{{\rm {eff}}}=\left(1-\frac{15}8g^2\right) \widetilde{g}_1=
\left(1-\frac{15}8g^2\right) \frac 52g_1=
\left(1-\frac{15}8g^2\right) \frac 52\ ,
\end{equation}
the desired perturbative result for the reduction of the $g_1$-factor due
to weak coupling to an $H_g$ mode. The factor $\frac 52$ reflects the $L=5$
parentage and $\left(1-\frac{15}8g^2\right) $ is the (weak-coupling) Ham
reduction factor\cite{Abragam} of this DJT problem, correctly coincident
with that obtained for a general vector observable by Bersuker and
Polinger.\cite{Bersuker} As anticipated, the reduction factor reflects the
increased ``effective mass'' of the $t_{1u}$ electron, as it carries along
some ionic mass while orbiting.

However, the coupling in C$_{60}$ is not really weak, and perturbation
theory is essentially only of qualitative value. For quantitative accuracy,
we can instead solve the problem by numerical (Lanczos)
diagonalization,\cite{AMT,Gunnarsson,Reno} which is feasible up to
realistically large coupling strengths. On a basis of states
\begin{equation}
\label{Basis}\Psi =\sum \epsilon _{k_1\mu _1,...k_N\mu _N,M,\sigma
}b_{k_1\mu _1}^{\dagger }...b_{k_N\mu _N}^{\dagger }c_{M\sigma }^{\dagger
}|0>
\end{equation}
(where $|0>$ is the state with no vibrons and no electrons), truncated to
include up to some maximum number $N$ of vibrons, ($N$ must be larger for
larger coupling) we diagonalize the Hamiltonian operator (\ref{Hschemat:eq}),
and take the numerical derivative of the ground-state energy with respect
to the magnetic field ${\cal B}$. Again, we consider here only the orbital
part, and ignore spin for the time being. In Fig.~\ref{Gfactor:fig} we plot
the resulting reduction of $g_1$-factor as a function of $g^2$ for a single
$H_g$ mode. The initial slope at $g=0$ coincides correctly with
$-\frac{15}8$, while at larger coupling, the behavior is compatible with
the expected Huang-Rhys--type decrease, $\sim \exp (-\chi g^2).$

We now repeat the same diagonalization including all the eight $H_g$
vibrons with their realistic couplings, as extracted by fitting gas-phase
photoemission spectra of C$_{60}^{-}$.\cite{Gunnarsson} Including, as in
our previous calculation of ground state and excitation
energies,\cite{Gunnarsson,Reno} up to $N$=5 vibrons for an accuracy of better
than two decimal figures, we obtain\cite{GfacScaling:nota} for the orbital
factor of the $t_{1u}$ LUMO of C$_{60}$ a final value of $0.17$,
whence
\begin{equation}
\label{Geff2}g_1^{{\rm {eff}}}=0.17\widetilde{g}_1=0.17\dot\frac 52
\simeq 0.43\ .
\end{equation}

With this orbital $g_L$-factor, we can now move on to compute the overall
$g_J$-factor in a realistic situation, where however spin-orbit must be
included.

\section{Spin-orbit coupling in the $t_{1u}$ LUMO, and results}
\label{so:sect}

The magnitude of the spin-orbit coupling $\lambda$ in the $t_{1u}$ state of
C$_{60}$ is not known. We estimate it by using straightforward
tight-binding, as follows. Starting from $2s$ and $2p_x,2p_y,2p_z$ orbitals
for each C atom, and including spin degeneracy, we diagonalize the 480x480
first-neighbor hopping Hamiltonian matrix to obtain all the molecular
orbitals.  Spin-orbit in this scheme is obtained by adding to the hopping
Hamiltonian a local coupling on each individual carbon in the form
\begin{equation}
H^{so}=\lambda_{\rm at} \sum_i {\bf L}_i\cdot {\bf S}_i \ .
\end{equation}
The level splitting introduced by this term defines the precise value of
spin-orbit coupling for each molecular orbital.  We are dealing with
$\pi$-states, which are unaffected by spin-orbit in a planar case, such as
in graphite. However, in fullerene, due to curvature, there will be an
effect.  In particular the splitting between the LUMO states
$^2{t_{1u}}_{\frac 12}$ and $^2{t_{1u}}_{\frac 32}$ gives the spin-orbit
coupling for the LUMO.  Our calculation yields $\lambda =0.9\cdot
10^{-2}\lambda_{\rm at}$ for the $t_{1u}$ state of C$_{60}$ when all bonds
are assumed to have equal lengths, slightly increasing to
\begin{equation}
\lambda =1.16\cdot 10^{-2}\lambda_{\rm at} \ ,
\end{equation}
when bond alternation is included.  Since $\lambda_{\rm at}=<2p_z\left|
{\bf L\cdot S}\right| 2p_x> \simeq 13.5$ cm$^{-1}$,\cite{Friedrich} we
conclude that the effective spin-orbit splitting of a $t_{1u}$ electron in
C$_{60}^{-}$ is of the order of $0.16$ cm$^{-1}=19\mu eV$. This value is
exceedingly small, due both to the small value of $\lambda$ in carbon, a
low-Z element, and (mainly) to the large curvature radius of C$_{60}$.  For
relatively large radius $R$, as appropriate to fullerenes and nanotubes,
one can expect a small spin-orbit effect in $\pi$-states, of order $\lambda
\sim 1/R^{2}$, the lowest power of curvature which is independent of its
{\em sign}. The
$\pi$-electron radius of C$_{60}$, $R \sim$ 5\AA, is one order of magnitude
larger than in the carbon atom, correctly suggesting a reduction of two
orders of magnitude from the atom to C$_{60}$.  Larger splittings of
30-50cm$^{-1}$ observed in luminescence spectra had earlier been attributed
to spin-orbit.\cite{Gasyna} These values are incompatible with our
estimate, and we conclude that these splittings must be of different
origin, unless an enhancement of two orders of magnitude over the gas
phase could somehow arise due to the host matrix.

We can now include this spin-orbit coupling in the calculation of the full
$g$-factor. A strong--spin-orbit approach\cite{ob92} $\lambda\gg E_{\rm
JT}$, relevant for some JT transition impurities, is not useful here,
since
clearly $\lambda \ll E_{\rm JT}\approx 140$meV\cite{Gunnarsson,Reno}
$ \sim
\hbar \omega $. In this case, the purely orbital description of the Berry
phase DJT of Refs.\ \cite{AMT,MTA} provides the correct gross features, to
which spin-orbit adds small splittings. These splittings are controlled by
the $g_J$ factors of Eq.~(\ref{CoupledGfac:eqn}). In our $t_{1u}\bigotimes
H_g$ case, an effective ``$\widetilde{L}=1$'' ground state is turned,
for positive $\lambda$, into a
``$J=\frac 12$'' ground-state doublet and a ''$J=\frac 32$'' excited
quartet. If we assume the usual weak-field limit $\mu_B B\ \ll\
\lambda$, we can recast Eq.~(\ref{CoupledGfac:eqn}) in
the form
\begin{equation}
\label{GJ1/2}
g_{\frac 12}=-\frac 13g_S+\frac 43g_L\simeq -\frac 23+\frac 43 g_L
\end{equation}
and
\begin{equation}
\label{GJ3/2}
g_{\frac 32}=\frac 13g_S+\frac 23g_L\simeq \frac 23+\frac 23 g_L\ .
\end{equation}
According to these linear formulae, the orbital reduction factor $g_1$
(Fig.~\ref{Gfactor:fig}) is easily deformed on the vertical axis to give
$g_J$. Since $g_L$ decreases from 1 to 0 for increasing JT coupling, we see
that while $g_{\frac 32}$ is always positive, $g_{\frac 12}$ may instead
become {\em negative} at large JT coupling.  For example, if the $L=5$
parentage enhancement is neglected, then $g_{\frac 12}$ ranges from $\frac
23$ at zero coupling to $-\frac 23$ at strong coupling.  Including the
$\frac 52$ orbital parentage factor, $g_{\frac 12}$ finally varies from
$\frac 83$ to $-\frac 23$.

The physical reason for a possible overall negative $g$-factor for $J=\frac
12$ at large e-v coupling, where $g_L \sim e^{-g^2}$, is also clear. The
$J=\frac 12$ overall{\em mechanical} angular momentum is dominated by $L=1$
orbital component. The {\em magnetic} moment is instead dominated by the spin
component, due to the strong reduction of the orbital part. But in the
$J=\frac 12$ state, spin and orbit are coupled (mainly) upside down, whence
the sign inversion.

Inserting the orbital $g_L$ factor (\ref{Geff2}) in
Eqs.~(\ref{GJ1/2},\ref{GJ3/2}), we get the final effective $g$-factor
for the ground state manifold of gas-phase C$_{60}^{-}$
\begin{equation}
\label{Geff3}
g_{\frac 12}^{{\rm {eff}}}=-0.1\ ,\ \ g_{\frac 32}^{{\rm {eff}}}=0.95\ .
\end{equation}

\section{Intermediate field}
\label{intfield:sect}

We have just obtained in Sect.\ \ref{so:sect} a slightly negative value for
the $g$-factor of the ``$J=\frac 12$'' ground state. This result, valid in
the approximation of $\mu_B B\ \ll\ \lambda \ \ll \ E_{\rm JT}$, becomes
unapplicable as soon as $\mu_B B\ \sim\ \lambda$. In an ideal EPR-like
experiment on gas-phase C$_{60}^-$, the resonant quantum $h\nu$ is easily
comparable with, or larger than, the spin-orbit frequency scale $\lambda/h
= 4.7$GHz. For example, standard X-band EPR employs a larger frequency of
9.5GHz, which resonates at $B=0.339$T, for a free spin.

For ease of comparison in Fig.~\ref{specEPR:fig} we report the full
spectrum of
the Zeeman-- and spin-orbit--split low-energy states, calculated
under the assumption that
$\mu_B B\ \ll\ E_{\rm JT}$, but for general spin-orbit strength,
and increasing magnetic
field.  Here, arrows indicate the symmetry-allowed microwave absorption
transitions,
matching an arbitrary excitation frequency 9.5GHz. EPR-like lines should
ideally appear at the
corresponding values of the field. Of the seven lines expected, two
correspond to (apparent) $g$-factors vastly larger than 2, and five to
$g$-factors
vastly smaller
than 2.  These $g$-factors are only apparent, since they depend on the
field (since it is not weak), and through it on the
frequency chosen. As the figure
indicates, the
weak-field limit for C$_{60}^{-}$ should only really be achieved with
fields in the order of a few hundred Gauss.

\section{Discussion and conclusions}
\label{discuss:sect}

Three main predictions result from the present calculation.  First, that
the molecular spin-orbit splitting is very small, a fraction of a degree
Kelvin.  Second, the reduced spin-orbit splitting is now wholly comparable
to typical Zeeman energies, for fields of a few KGauss. Hence, a strong and
uncommon field-dependence of the $g$-factors is predicted.
Third, we find that the orbital gyromagnetic factor is strongly
reduced by vibron coupling, and so therefore are the effective weak-field
$g$-factors for all the low-lying states. In particular, the isolated
C$_{60}^{-}$ ion in its ``$J=1/2$'' ground state should be essentially non
magnetic, in fact slightly diamagnetic, if cooled below $T\approx
\frac{E_{3/2}-E_{1/2}}{k_B} \sim 0.2$K.

For C$_{60}^{-}$ in solid ionic salts at 77K and room temperature, Kato
{\it et al.}\cite{kato} have found $g^{{\rm {eff}}}\simeq 1.999$, with
slight anisotropies due to the lattice. Similar $g^{{\rm {eff}}}$values are
also obtained for C$_{60}^{-}$ in molecular sieves\cite{Keizer}, as well as
in various solvents and salts\cite{Penicaud,Schell-Sorokin,Khaled,Boyd}.
These $g$-factors are relatively close to the bare-spin value, implying
that the dynamical orbital effects discussed above are apparently quenched by
coupling to the matrix. In the photoexcited triplet state of neutral
C$_{60}$, some evidence has been found for nonthermal jumps between
JT valleys\cite{mehring}, but apparently none for orbital magnetism.
In a majority of these systems, all goes as if the extra electron
of C$_{60}^{-}$, or the extra electron-hole triplet pair of C$_{60}^{-}$
occupied a nondegenerate level, as expected in the
static JT case. A detailed discussion of the quenching of quantum
orbital effects is beyond the scope of this work. However, we think
that coupling to the host matrix, however weak it may be in some cases,
must be responsible for the apparent quenching from DJT to static JT.
One possibility, for example, is that the extra electron on  C$_{60}^{-}$
acts to strongly polarizes the surrounding, which in turn slows down and damps
the quantum mechanical electron tunneling between different JT valleys.
Insofar as these couplings seem ubiquitous and fatal to the DJT,
we would provisionally tend to conclude that gas-phase studies may
represent the only serious possibility for the observation of orbital moments
in fullerene ions.

Stern-Gerlach--like measurements of magnetic moment in
gas-phase C$_{60}^{-}$, or other similar experiments, are
therefore called for to provide a definitive confirmation of the striking
quantum orbital effects described in this paper. Since to our knowledge
this would be new, we feel that such experiments
should be considered, even if very difficult to carry out.

Gas-phase measurements would presumably be easier in the $^3t_{1g}$
triplet exciton
state of {\em neutral} C$_{60}$.  This state has a number of
similarities\cite{triplet:note} to that of C$_{60}^{-}$ which we have just
described.  However, here $S=1$ and $\widetilde{L}=1$, leading to a $J=0$
singlet ground state\cite{singlet:note} for the triplet exciton with DJT
and spin-orbit coupling, and the magnetic anomalies will have to be sought
in the lowest excited states.

We acknowledge support from NATO through CRG 92 08 28, and the EEC, through
Contracts ERBCHRXCT 920062, and ERBCHRXCT 940438.


\begin{thebibliography}{10}
\bibitem{triplet:note}
{ The $^3t_{2g}$ electronic configuration of the lowest exciton state, from the
  point of view of orbital symmetry, is analogous to the $^2t_{1u}$
  configuration of C$_{60}^-$. As a consequence, it also undergoes a JT effect
  associated to coupling with the $H_g$ modes. The $^3t_{2g}\otimes H_g$ DJT
  problem is also endowed with a Berry phase, similar to the $^2t_{1u}\otimes
  H_g$}.
\bibitem{GfacScaling:nota}
{ The DJT reduction factor is obtained from extrapolation of its finite-cutoff
  values 0.20993 ($N=3$), 0.18257 ($N=4$), 0.17393 ($N=5$), 0.17167 ($N=6$)}.
\bibitem{singlet:note}
{ This is the order of levels if the sign of the spin-orbit parameter for the
  $^3t_{2g}$ state is positive. This sign is not obvious. For instance, we also
  find $\lambda/\lambda_{\rm at} \sim 0.01$ for the $h_u$ HOMO, but
  $\lambda/\lambda_{\rm at} \sim -0.01$ for the $t_{1g}$ LUMO+1. For the
  $^3t_{2g}$ exciton the spin-orbit sign is probably positive, as for the HOMO
  and the LUMO parent orbitals}.
\end{thebibliography}

\begin{figure}
\caption{The reduction factor of the orbital magnetic moment and $g$-factor
        $g_1$, due to DJT coupling of the $t_{1u}$ state to a single $H_g$
        vibron. The calculation is done by exact diagonalization, the
        truncated basis set including up to $N=11$ vibrons.  Note the fast
        linear decrease at small $g$, and also the $e^{-g^2}$ decay as
        expected at large $g$.  In C$_{60}^{-}$, including eight coupled
        $H_g$ modes instead of one, the overall reduction factor
        obtained with a similar calculation and realistic couplings is 0.17
        (see sect. III).
\label{Gfactor:fig}}
\end{figure}

\begin{figure}
\caption{The low-energy levels of C$_{60}^{-}$ ($g_1^{\rm eff}=0.43$),
        calculated for increasing magnetic field $B$. All energies,
        including $\mu_B B$, are measured in units of the molecular
        spin-orbit coupling $\lambda$, which we estimate to be about
        4.7GHz$\cdot h$
	(Sect.~\protect\ref{so:sect}).  The vertical arrows indicate
	allowed microwave absorption transitions, for a frequency of
	9.5GHz. The apparent $g$-factor values corresponding to these
        transitions differ vastly from the free spin value, and are heavily
        dependent  upon the DJT coupling parameters, and the relative
        value of spin-orbit coupling. The weak-field region shows clearly
        the $J=\frac 12$ structure (inset), exhibiting the weak negative
        $g$-factor.  Note also the two ground-state level crossings,
        corresponding to zero-temperature magnetization jumps of
        0.14$\mu_B$ and 0.56$\mu_B$, for field values of 0.23T and 0.57T
        respectively.
\label{specEPR:fig}}
\end{figure}

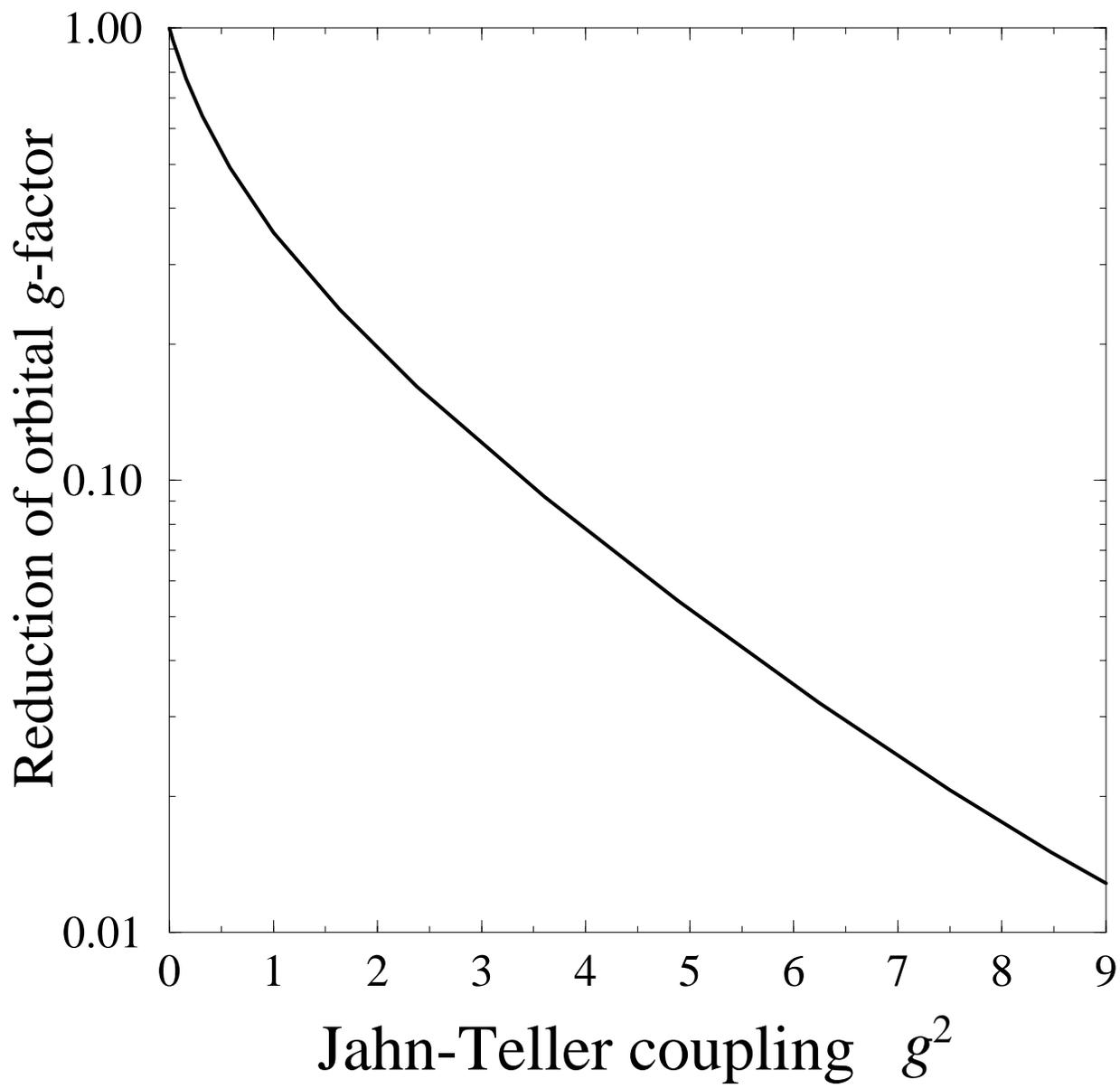
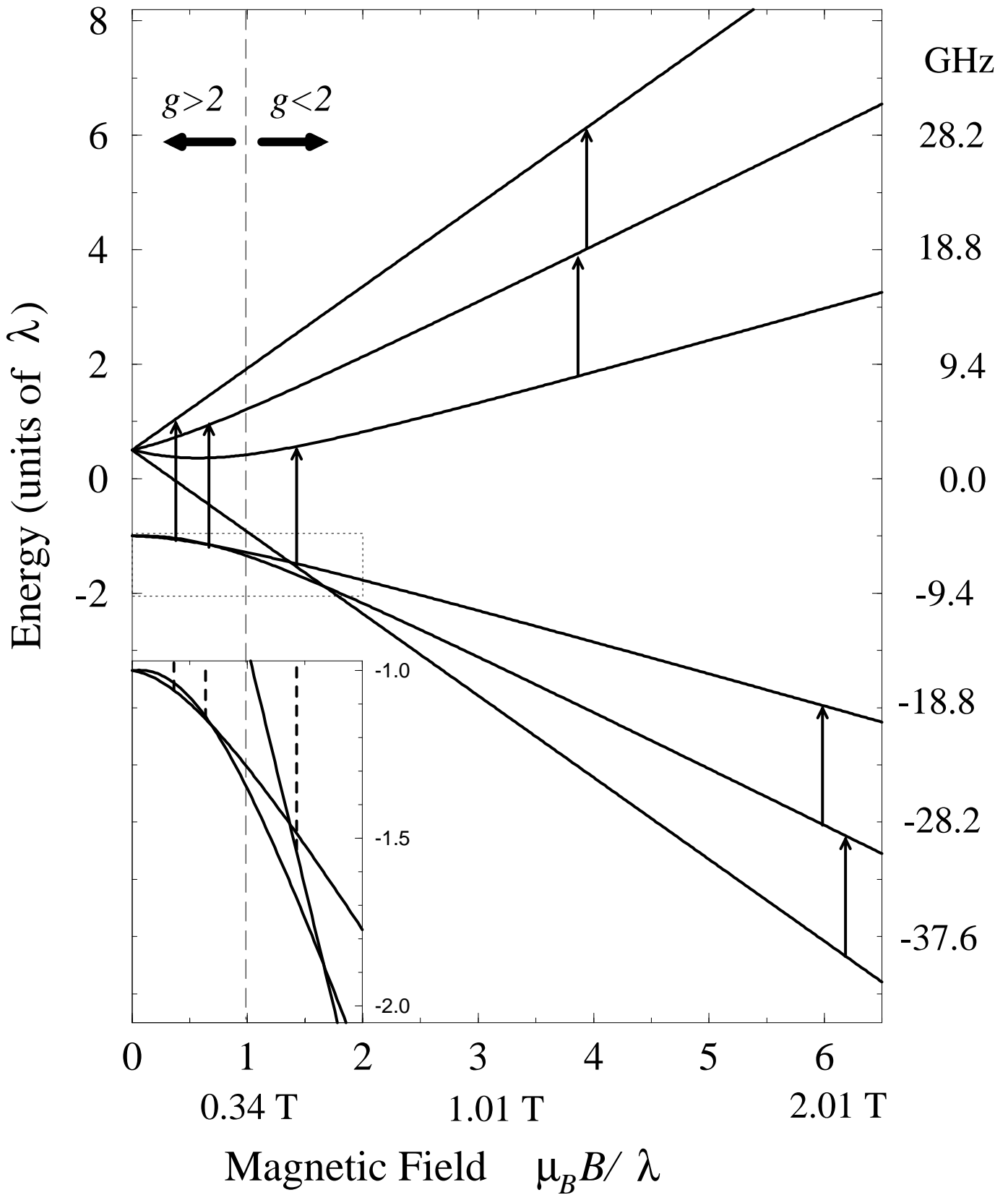
\end{document}